\documentclass[twocolumn,showpacs,prl]{revtex4}
\usepackage{amsmath}
\usepackage{graphicx}
\usepackage{dcolumn}
\usepackage{bm}
\usepackage{amsfonts}
\usepackage{amssymb}
\begin{document}
\title{ Microscopic theory of thermal phase slips in clean narrow superconducting wires }
\author{A. Zharov$^{1,2}$, A. Lopatin$^{1}$, A. E. Koshelev$^{1}$, and V.M. Vinokur$^{1}$ }
\affiliation{$^{1}$Materials Science Division, Argonne National
Laboratory, Argonne, Illinois 60439, USA\\ $^2$Illinois Institute of
Technology,  3300 S. Federal St. Chicago, IL 60616, USA}
\date{\today}

\begin{abstract}
We consider structure of a thermal phase-slip center for a simple
microscopic model of a clean one-dimensional superconductors in
which superconductivity occurs only within one conducting channel or
several identical channels. Surprisingly, the Eilenberger equations
describing the saddle-point configuration allow for exact analytical
solution in the whole temperature and current range. This solution
allows us to derive a closed expression for the free-energy barrier,
which we use to compute its temperature and current dependences.
\end{abstract}
\maketitle

%\section{Introduction}
%
%%%%%%%%%%%%%%%%%%%%%%%
%%%%%%%%%%%%%%%%%%%%%%%
Recent advance in fabrication techniques brought ultrathin
superconductor wires into active experimental investigation
\cite{Bezryadin,Zgirski,Tian,Altomare} as they may play an important
role in future electronics. These studies brought about the
fundamental question of the limits till which thin wires retain a
key property of superconductors, the ability to carry currents
without dissipation. It was well established that in extremely
narrow superconductors thermal fluctuations can occasionally drive a
small part of the wire into a normal state. When there is only one
channel for a supercurrent to pass, these fluctuations interrupt its
flow and cause a finite effective resistance. This phenomenon has
been understood by Langer and Ambegaokar (LA) as thermally activated
\textit{phase slips} ~\cite{Langer}: local abrupt switches of the
system between its metastable states in which events phase jumps
discontinuously by $\pm 2\pi$, see also Ref.\ \onlinecite{Tinkham}
for a review. The corresponding resistance exhibits Arrhenius
law-like temperature behavior, $\rho\propto\exp(-\delta F/T)$, where
$\delta F$ is the energy barrier separating these states. In a
phase-slip event the order parameter goes through the saddle-point
configuration in which it changes sign as function of coordinate and
vanishes at one point. The barrier for the phase slip, $\delta F$,
is given by the energy of this saddle-point configuration. Later
preexponential factor in resistivity has been calculated by McCumber
and Halperin \cite{Halperin} and Langer-Ambegoakar-McCumber-Halperin
(LAMH) theory has been confirmed experimentally on tin
whiskers~\cite{FirstExpConf}. The dynamical theory of phase slips at
large currents has been further extended in Refs.\
\cite{Ivlev,Peeters}.

The mechanism of appearance of a finite resistivity in a thin wire
can be described as follows~\cite{Tinkham}. A finite voltage drop
across the wire means that the order parameter phase difference
$\phi_{12}$ between the ends of the wire has to increase linearly
with time. At first glance, this should inevitably lead to
continuous increase of the supercurrent flowing through the wire,
which is obviously is inconsistent with steady state. The resolution
of this apparent inconsistency is that the current relaxes via the
phase-slip events, in which the order parameter vanishes at some
point in the superconductor, allowing a phase difference between
opposite sides of this point to change by $2\pi$. As a result of
this event, a phase jump equal to $2\pi$ appears at one point of
superconductor reducing the phase gradient in the rest part of the
wire. To maintain a steady state, such events must occur with the
average frequency $2e\langle{V}\rangle/\hbar$.

Phase slip events can be caused also by quantum tunneling, which at
sufficiently low temperatures can prevail over the thermal
activation, and this possibility is being widely searched and
debated~\cite{Bezryadin,Zgirski,Tian,Altomare,Zaikin}. Yet, the role
and the very fact of observation of quantum phase slips remain a
controversy. It seems now plausible that thermally activated phase
slips may dominate down to very low temperatures and that quantum
fluctuations do not add any significant corrections to the
resistance in almost whole experimentally accessible temperature
range.  As original LAMH theory  \cite{Langer,Halperin} was
developed within the framework of the Ginzburg-Landau (GL) approach,
it holds, strictly speaking, only in the temperature interval near
the critical temperature, $T_{c}$, of the superconductor transition.
On the other hand, in sufficiently narrow wires a finite
thermally-activated resistivity can be observed down to temperatures
which are significantly lower than $T_{c}$,  There is thus an
experimental quest for extension of the LAMH theory to low
temperatures, where the phase slips have to be analyzed within the
microscopic theory.

In this Letter we investigate a simple model of a clean
one-dimensional superconductor containing either only one conducting
channel or a few degenerate channels. It turns out that this model
allows for full analytical treatment of a free-energy saddle point
configuration. An exact analytical solution of Eilenberger equation
is found, both at zero and finite supercurrent. This solution allows
us to derive a closed expression for the temperature and current
dependent free-energy barrier.

%\section{Derivation of a free-energy saddle-point solution}
%
We consider the simplest case of a narrow clean superconducting wire
containing either only one superconducting channel or several
degenerate channels, i.e. with the same Fermi velocity, $v_{F}$. The
wire diameter is assumed to be smaller than the London penetration
depth; one can thus neglect the magnetic field generated by the
currents. We will find an order parameter configuration which
corresponds to free-energy saddle point in such a wire. As the
saddle-point configuration is a stationary point in the parameter
space, it is described by the same equations as the equilibrium
state. In the quasiclassical approximation such superconducting wire
is described by the gap parameter $\Delta(x)$ and the Green's
functions, $g,f,f^{\dag}$ depending on the coordinate along the wire
$x$, the Matsubara frequencies $\omega_{n}\!=\!2\pi T(n\!+\!1/2)$
and the location on the Fermi surface, which is reduced to two
points corresponding to forward and back motion. The Green's
functions are connected by the normalization condition
$g^{2}\!+\!ff^{\dag}\!=\!1$ and obey the one-dimensional Eilenberger
equations \cite{eilenberger}:
\begin{subequations}
\begin{align}
v_{F}\frac{\partial g}{\partial x}+\Delta f^{\dag}-\Delta^{\ast}f &
=0,\label{ei1re}\\
-v_{F}\frac{\partial f}{\partial x}-2\omega_{n}f+2\Delta g &  =0,\\
v_{F}\frac{\partial f^{\dag}}{\partial
x}-2\omega_{n}f^{\dag}+2\Delta^{\ast}g &  =0,
\end{align}
\end{subequations}
with $v_{F}=\pm|v_{F}|$. It is convenient to introduce new
dimensionless variables
\[
x\rightarrow x/\xi,\ \Delta\rightarrow \Delta/T_{c},\ T\rightarrow
T/T_{c},
\]
where $\xi=v_{F}/T_{c}$ is the coherence length and $T_{c}$ is the
critical temperature. In new variables the Eilenberger equations for
forward direction, $v_{F}>0,$ become
\begin{subequations}
\begin{align}
\frac{\partial g}{\partial x}+\Delta f^{\dag}-\Delta^{\ast}f &
=0,\label{ei2re}\\
-\frac{\partial f}{\partial x}-2\omega_{n}f+2\Delta g &  =0,\\
\frac{\partial f^{\dag}}{\partial
x}-2\omega_{n}f^{\dag}+2\Delta^{\ast}g & =0.\label{ei3re}
\end{align}
These equations must be completed with the self-consistency
equation,
\end{subequations}
\begin{equation}
\frac{\Delta(x)}{\lambda}=\pi T\sum_{\omega_{n}}\langle f(\omega_{n}
,x,\hat{k}_{F})\rangle_{F},\label{self}
\end{equation}
and the formula for the supercurrent,
\begin{equation}
j=-2\pi ieN(0)T\sum_{\omega_{n}}\langle v_{F}g(\omega_{n},x,\hat{k}
_{F})\rangle_{F},\label{superc}
\end{equation}
where $N(0)=1/(\pi v_{F})$ is the density of states. The summation
is going over all Matsubara frequencies $n=-\infty \ldots \infty$.
For the one-dimensional problem averaging over the Fermi surface,
$\langle\ldots\rangle_{F}$, reduces to summation over two points
\begin{align*}
\langle f(\omega_{n},x,\hat{k}_{F})\rangle_{F} &
\!=\!\frac{1}{2}\left(
f(\omega_{n},x,k_{F})\!+\!f(\omega_{n},x,-k_{F})\right),  \\
\langle v_{F}g(\omega_{n},x,\hat{k}_{F})\rangle_{F} &
\!=\!\frac{v_{F}}{2}\left(
g(\omega_{n},x,k_{F})\!-\!g(\omega_{n},x,-k_{F})\right).
\end{align*}
The Green's functions in the opposite directions are connected by
relations $f(-k_{F})=\left[  f^{\dag}(k_{F})\right]  ^{\ast}$,
$f^{\dag}(-k_{F})=\left[ f(k_{F})\right]  ^{\ast}$ and
$g(-k_{F})=\left[  g(k_{F})\right]  ^{\ast}$.

Let a finite current flow through the wire. Then the order
parameter, $\Delta$, and the anomalous Green functions $f$ and
$f^{\dag}$ can be represented in the form of plane waves with slowly
varying in space amplitudes, $\Delta,f\propto
e^{ikx};\;\;\Delta^{\ast},f^{\dag}\propto e^{-ikx}$, where the wave
vector $k$ is determined by the supercurrent.  Defining new
variables $f_{\pm}=\left(  f\pm f^{\dag}\right) /2$, one rewrites
Eqs.\ (\ref{ei2re}-\ref{ei3re}) as
\begin{subequations}
\label{ei3m}
\begin{align}
\frac{\partial g}{\partial x}+2i\Delta_{I}f_{+}-2\Delta_{R}f_{-} &
=0,\label{ei3mg}\\
\frac{\partial f_{-}}{\partial
x}-2\Delta_{R}g+2{\omega}_{n}^{\prime}f_{+} &
=0,\label{ei3mf-}\\
\frac{\partial f_{+}}{\partial
x}-2i\Delta_{I}g+2{\omega}_{n}^{\prime}f_{-} & =0,\label{ei3mf+}
\end{align}
\end{subequations}
with $\Delta_{R}=\text{Re}\Delta$, $\Delta_{I}=\text{Im}\Delta$, and
${\omega }_{n}^{\prime}=\omega_{n}+ik/2$. Using relations between
the Green's functions for the opposite directions, we can rewrite
the self-consistency condition (\ref{self}) separately for the real
and imaginary parts of $\Delta$ as
\begin{subequations}
\label{selfBoth}
\begin{align}
\frac{\Delta_{R}(x)}{\lambda} &  =\pi T\sum_{\omega_{n}}\text{Re}
f_{+}(\omega_{n},x),\label{self1}\\
\frac{\Delta_{I}(x)}{\lambda} &  =\pi T\sum_{\omega_{n}}\text{Im}
f_{-}(\omega_{n},x).\label{self2}
\end{align}
\end{subequations}
Homogeneous solution for superconductor in equilibrium is well
known,
\begin{align}
&  f_{+}^{(0)}=S_{n}\Delta_{R0},\ f_{-}^{(0)}=iS_{n}\Delta_{I},\
g^{(0)}
=S_{n}{\omega}_{n}^{\prime}\label{HomogGreen}\\
&  \text{with
}S_{n}=1/\sqrt{{\omega_{n}^{\prime}}^{2}+|\Delta_{0}|^{2}
}.\label{DefS}
\end{align}
The absolute value of the equilibrium order parameter and
supercurrent can be found from Eqs.\ (\ref{self}) and (\ref{superc})
as
\begin{align}
&  \pi T\sum_{\omega_{n}}\left(  \text{Re}\frac{1}{\sqrt{{\omega
_{n}^{\prime2}}+|\Delta_{0}|^{2}}}-\frac{1}{|\omega_{n}|}\right)
=\ln
T, \label{delq}\\
&  j=-2eT\sum_{\omega_{n}}\text{Im}\frac{\omega_{n}^{\prime}}{\sqrt
{{\omega_{n}^{\prime}}^{2}+|\Delta_{0}|^{2}}}.\label{superc2}
\end{align}
Note that the depairing action of superfluid velocity in our model
is equivalent to the depairing due to the Zeeman splitting, see,
e.g., Ref.\ \onlinecite{maki}. As a consequence, in our system the
temperature dependence of the critical current is nonmonotonic,
similar to the Zeeman-effect upper critical field. Such temperature
dependence is a peculiar property of clean one-dimensional
superconductors and it is very different from higher dimensional
systems.

We will demonstrate now that Eqs.\ (\ref{ei3m}) and (\ref{selfBoth})
allow for exact analytical solution. First, from Eqs.\ (\ref{ei3m})
we obtain a useful relation between the Green's functions
\begin{equation}
i\Delta_{I}\frac{\partial f_{-}}{\partial
x}-\Delta_{R}\frac{\partial f_{+} }{\partial
x}-{\omega}_{n}^{\prime}\frac{\partial g}{\partial x}=0\label{int1}
\end{equation}
In general, we cannot reduce this relation to the first integral of
equations (\ref{ei3m}a-c). To proceed, we make some strong
assumptions about the form of the order parameter and Green's
functions. We will check consistency of these assumptions from the
obtained solution. Namely, we assume that (i) the imaginary part of
the order parameter $\Delta_{I}$ is coordinate independent and (ii)
$f_{+}$ can be found in the variable-separated form
\begin{equation}
f_{+}(\omega_{n},x)=S_{n}\Delta_{R}(x),\label{VarSep}
\end{equation}
where the frequency term $S_{n}$ is assumed to be coordinate
independent and therefore it has to be given by Eq.\ (\ref{DefS}).
In this case we can rewrite Eq.\ (\ref{int1}) as a first integral,
\[
\frac{\partial}{\partial x}\left[
i\Delta_{I}f_{-}-S_{n}\frac{\Delta_{R}^{2}
}{2}-{\omega}_{n}^{\prime}g\right]  =0,
\]
which gives
$i\Delta_{I}f_{-}-S_{n}\frac{\Delta_{R}^{2}}{2}-{\omega}_{n}^{\prime
}g=\textmd{C}. $ The constant $\textmd{C}$ can be found using
asymptotics far away from the phase-slip center. Substituting for
Green's functions their equilibrium values from Eq.
(\ref{HomogGreen}), we obtain $\textmd{C} =-S_{n}\left(
{\omega_{n}^{\prime}}^{2}+\Delta_{I}^{2}+\Delta_{R0} ^{2}/2\right)
$, which gives
\begin{equation}
i\Delta_{I}f_{-}-{\omega}_{n}^{\prime}g=-S_{n}\left(
{\omega_{n}^{\prime}
}^{2}+\Delta_{I}^{2}+\frac{\Delta_{R0}^{2}-\Delta_{R}^{2}}{2}\right).
\label{Relation}
\end{equation}
Taking derivative of Eq. (\ref{ei3mf+}) and excluding $\partial
g/\partial x$ and $\partial f_{-}/\partial x$ from Eqs.
(\ref{ei3mg}) and (\ref{ei3mf-}), we obtain
\begin{equation}
\frac{\partial^{2}f_{+}}{\partial x^{2}}-4\left(
\Delta_{I}^{2}+{\omega} _{n}^{\prime2}\right)
f_{+}-4\Delta_{R}\left(  i\Delta_{I}f_{-}-{\omega}
_{n}^{\prime}g\right)  =0.
\end{equation}
Substituting for $f_{+}$ the presentation (\ref{VarSep}) and using
the relation (\ref{Relation}), we see that the frequency part indeed
drops out and this equation reduces to the differential equation for
$\Delta_R$,
\begin{equation}
\frac{\partial^{2}\Delta_{R}}{\partial x^{2}}+2\Delta_{R0}^{2}\Delta
_{R}-2\Delta_{R}^{3}=0,
\end{equation}
which essentially coincides with the Ginzburg-Landau equation.
However, in our case it describes the phase-slip center in the whole
temperature range. This equation has a kink soliton-like solution
\begin{equation}
\Delta_{R}(x)=\Delta_{R0}\tanh\left(  \Delta_{R0}x\right)  .
\label{OrderParSol}
\end{equation}
Corresponding Green's functions are
\begin{subequations}
\label{GreenSol}
\begin{align}
f_{+} &  =S_{n}\Delta_{R0}\tanh\left(  \Delta_{R0}x\right)  ,\label{gsol}\\
f_{-} &  =S_{n}\left(  i\Delta_{I}-\frac{\Delta_{R0}^{2}}{2({\omega
_{n}^{\prime}}+i\Delta_{I})}\frac{1}{\cosh^{2}\left(
\Delta_{R0}x\right)
}\right)  ,\\
g &  =S_{n}\left(
{\omega_{n}^{\prime}}+\frac{\Delta_{R0}^{2}}{2({\omega}
_{n}^{\prime}+i\Delta_{I})}\frac{1}{\cosh^{2}\left(
\Delta_{R0}x\right) }\right)  .
\end{align}
\end{subequations}
One can verify that the normalization constraint
$g^{2}+f_{+}^{2}-f_{-}^{2}=1$ is satisfied. The assumed coordinate
independence of the imaginary part of $\Delta$ requires that the
coordinate-dependent part of Eq. (\ref{self2}) vanishes. This gives
the following equation
\begin{equation}
\sum_{\omega_{n}}\text{Im}\left[
\frac{1}{{\omega_{n}^{\prime}}+i\Delta_{I}
}\frac{1}{\sqrt{{\omega_{n}^{\prime}}^{2}+|\Delta_{0}|^{2}}}\right]
=0.\label{EqDeltaI}
\end{equation}
which can be used to find $\Delta_{I}$, while $|\Delta_{0}|$ is
determined by Eq.\ (\ref{delq}) and
$\Delta_{R0}=\sqrt{|\Delta_{0}|^{2}-\Delta_{I}^{2}}$. Therefore, the
structure of the phase-slip center is fully defined.

%\section{Free energy barrier}
%
The probability of a phase slip event driven by thermal fluctuation
is proportional to the exponent $\exp\left(  -\delta F/ T\right) $,
where $\delta F$ is the free-energy barrier between homogeneous and
saddle-point state and $T$ is the temperature. Therefore, the next
point is to find the free-energy barrier.

The formula expressing the free energy in terms of the
quasiclassical Green's functions has been derived by Eilenberger in
his original paper \cite{eilenberger}. This formula can be
straightforwardly generalized to the one-dimensional case and in
real units it can be written as
\begin{align}
\label{freeenrgy0} &F\!=\!N(0)\!\int\! dx
\left(\frac{|\Delta|^{2}}{\lambda}- \pi
T\sum_{\omega_{n}}\left\langle \frac{v_{F}g}{2}\left(
\frac{\partial_{x} f}{f}\!-\!\frac{\partial_{x} f^{\dag}
}{f^{\dag}}\right)\right. \right.
\nonumber\\
&  + \Delta f^{\dag}+\Delta^{*} f+2\omega_{n} (g-1) \Big\rangle _{F}
\Big).
\end{align}
The phase-slip barrier is given by the energy difference between
homogeneous and inhomogeneous states, $\delta F=F-F_{\mathrm{hom}}$,
and the free-energy density of a homogeneous state is known to be
\begin{align}
\label{hom}\mathcal{F}_{\mathrm{hom}}= |\Delta|^{2}/\lambda-\pi
T\sum_{\omega_{n} }\left(
\sqrt{\omega_{n}^{2}+|\Delta|^{2}}-|\omega_{n}|\right).
\end{align}
Using the Eilenberger equations and constraint $g^{2}+ff^{\dag}=1$,
we can simplify the expression for free energy (\ref{freeenrgy0})
and rewrite it without space derivatives. With the self consistency
equation we can get rid of the coupling constant $\lambda$.
Subtracting the homogeneous-state free energy from the full
expression and using dimensionless units, we obtain the free energy
in units of $T_{c}$,
\begin{align}
\label{dfreeenrgy1}\delta F=\int\delta\mathcal{F} dx
\end{align}
with
\begin{align}
\label{dfreeenrgy}
&\delta\mathcal{F}=\pi^{-1}(|\Delta|^{2}-|\Delta_{0}|^{2})\ln
T+\nonumber\\
&T\text{Re}\sum_{\omega_{n}} \left[  \frac{|\Delta |^{2}-|\Delta_{0}
|^{2}}{|\omega_{n}|}- \left(  \frac{\Delta}{f}+\frac
{\Delta^{*}}{f^{\dag}} -\frac{2}{S_{n}} \right)  \right].
\end{align}
Using our saddle-point solution (\ref{OrderParSol}) and
(\ref{GreenSol}) and performing integration over $x$, we obtain for
$\delta F$,
\begin{align}
\delta F  & =T\text{Re}\sum_{\omega_{n}}\!\left(\! \ln
\frac{|\Delta_{0}|^{2}\!-i\Delta_{I}\omega_{n}^{\prime}+\Delta_{R0}\sqrt
{{\omega_{n}^{\prime}}^{2}+|\Delta_{0}|^{2}}}{|\Delta_{0}|^{2}\!-i\Delta
_{I}\omega_{n}^{\prime}-\Delta_{R0}\sqrt{{\omega_{n}^{\prime}}^{2}+|\Delta
_{0}|^{2}}}\right.  \nonumber\\
& \left.
-\frac{2\Delta_{R0}}{\sqrt{{\omega_{n}^{\prime2}}+|\Delta_{0}|^{2}}
}\right) \label{dfreeenrgy2}
\end{align}
This formula gives the barrier in terms of the sum over Matsubara
frequencies, which allows for analytical calculation in the limits
$T\rightarrow 0,1$ and can be easily computed numerically in general
case.

At zero temperature sum over Matsubara frequencies in
(\ref{dfreeenrgy2}) becomes an integral $T \sum_{\omega_{n}}
\rightarrow\int d\omega/2\pi $. At $T\rightarrow 0$ the gap
parameter $|\Delta_0|$ is given by the BCS value,
$\Delta(0)=(\pi/\gamma)T_c$ with $\gamma=1.7811$ and in our clean
case it remains current-independent up to the critical current
corresponding to $k\approx0.83$. Eq.\ (\ref{delq}) gives
$\Delta_{I}\rightarrow -k/2$ for $T\rightarrow 0$. Integration over
$\omega$ in Eq.\ (\ref{dfreeenrgy2}) can be evaluated and leads to a
very simple exact result for the zero-temperature barrier
\begin{align}
\delta
F(0,k)=\frac{2}{\pi}\Delta_{R0}=\frac{2}{\pi}\Delta(0)\sqrt{1-\left(
\frac{k}{2}\right)  ^{2}}
\end{align}
for $k<0.83$
\begin{figure}[ptb]
\begin{center}
\includegraphics[width=3.in]{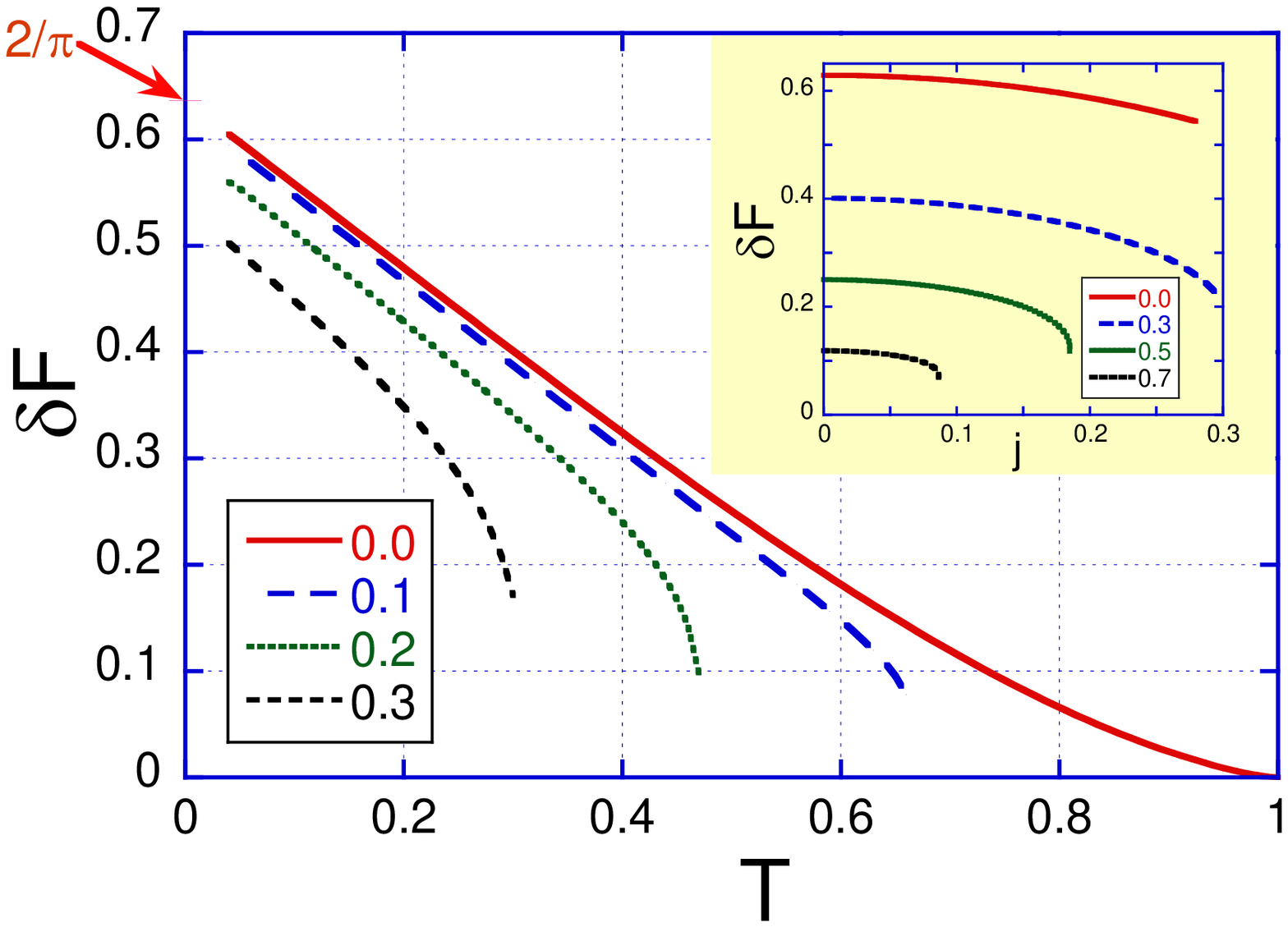}
\end{center}
\caption{ The temperature dependences of the phase-slip barrier in
units of $\Delta(0)$ for different currents. The legend indicates
values of supercurrent measured in units of $j_{0}=2 eT_{c}$. Note
that at finite current the barrier always jumps to zero from a
finite value. Inset shows current dependences of the barrier at
different temperatures.} \label{Fig-dF_T}
\end{figure}

We also can find the asymptotics of the barrier in the
Ginzburg-Landau region, $|T-T_{c}|\ll1$, where our results have to
be similar to the LAMH solution. In this temperature range
$\Delta\ll1$ and we can take into account only the first
non-vanishing term of the Taylor expansion of Eq.\
(\ref{dfreeenrgy2}). In this limit for a wire without current we
have
\begin{align}
\delta F=C_{0}\Delta(0)\left( 1-T\right) ^{3/2}, \label{BarrierGL}
\end{align}
where $C_{0}=\frac{8\sqrt{2}}{3\sqrt{7\zeta(3)}}\frac{\gamma}{\pi}
\approx0.737$ with $\zeta(z)=\sum_{n=1}^{\infty}n^{-z}$ is the
Riemann $\zeta $-function, $\zeta(3)\approx 1.202$.

For arbitrary temperatures and currents the free-energy barrier
cannot be represented in a simple form and has be computed
numerically. Figure \ref{Fig-dF_T} shows the temperature dependences
of the barrier in units of $\Delta (0)$ for different dimensionless
currents in units of $j_{0}=2eT_{c}$. We also show in the inset the
current dependences of $\delta F$ at several temperatures. One can
see that the temperature dependences of the barrier have almost
linear part at $T\rightarrow 0$. This is a consequence of the
superconducting gap vanishing at one point. It is well known that in
fully-gapped superconductors the temperature dependences of all
parameters are exponential. Accidently, the Ginzburg-Landau
asymptotics (\ref{BarrierGL}) actually provides a reasonable
approximation for the exact barrier in the whole temperature range.
Even at $T\rightarrow0$ the GL result exceeds the exact one only by
15\%.

Our results can be straightforwardly generalized for a case of
several identical conducting channels. If we have $M$ channels with
the same Fermi velocity, then the corresponding free-energy barrier
is just proportional to the number of channels, $\delta F_{M} =
M\delta F$. The current scale in Fig. \ref{Fig-dF_T} also scales
with the number of channels, $j_{0}\rightarrow j_{M}=2MeT_{c}$. It
is clear that the thicker the wire is and the more conducting
channels it has, the higher the free-energy barrier is.

In conclusion, we considered the model of thermally activated phase
slips in a clean one-dimensional superconductor with only one
conducting channel and found the exact analytical solution of
Eilenberger equations corresponding to the free-energy saddle point.
This solution allowed us to calculate the temperature and current
dependences of the free-energy barrier for thermal phase slips.

The authors would like to thank I.\ Beloborodov for critical reading
of the manuscript and useful comments. This work was supported by
the U.\ S.\ DOE, Office of Science, under contract \#
DE-AC02-06CH11357. 

\end{document}